\documentclass{aastex}
\usepackage{emulateapj5}
\usepackage{apjfonts}

\newcommand{\swas}{{\it SWAS}}
\newcommand{\hii}{H~{\sc ii}}
\newcommand{\water}{H$_2$O}
\newcommand{\wateriso}{H$_2^{18}$O}
\newcommand{\ceio}{C$^{18}$O}
\newcommand{\thco}{$^{13}$CO}
\newcommand{\chhhcch}{CH$_3$CCH}
\newcommand{\hh}{H$_2$}

\journalinfo{{\rm Published in} The Astrophysical Journal, 539:L97--L100,
             {\rm 2000 August 20}}
\slugcomment{Received 2000 January 28; accepted 2000 June 20; published
             2000 August 16}

\shortauthors{DISTRIBUTION OF WATER EMISSION IN M17SW}
\shorttitle{SNELL ET AL.}

\begin{document}

\title{The Distribution of Water Emission in M17SW}

\author{R. L. Snell\altaffilmark{1},
J. E. Howe\altaffilmark{1},
M. L. N. Ashby\altaffilmark{2},
E. A. Bergin\altaffilmark{2},
G. Chin\altaffilmark{3},
N. R. Erickson\altaffilmark{1},
P. F. Goldsmith\altaffilmark{4},
M. Harwit\altaffilmark{5},
S. C. Kleiner\altaffilmark{2},
D. G. Koch\altaffilmark{6},
D. A. Neufeld\altaffilmark{7},
B. M. Patten\altaffilmark{2},
R. Plume\altaffilmark{2},
R. Schieder\altaffilmark{8},
J. R. Stauffer\altaffilmark{2},
V. Tolls\altaffilmark{2},
Z. Wang\altaffilmark{2},
G. Winnewisser\altaffilmark{8},
Y. F. Zhang\altaffilmark{2},\\
and G. J. Melnick\altaffilmark{2}}

\altaffiltext{1}{Department of Astronomy, University of Massachusetts,
                 Amherst, MA 01003}
\altaffiltext{2}{Harvard-Smithsonian Center for Astrophysics, 60 Garden Street,
                 Cambridge, MA 02138}
\altaffiltext{3}{NASA Goddard Spaceflight Center, Greenbelt, MD 20771}
\altaffiltext{4}{National Astronomy and Ionosphere Center, Department of
                 Astronomy, Cornell University, Space Sciences Building,
                 Ithaca, NY 14853-6801}
\altaffiltext{5}{511 H Street SW, Washington, DC 20024-2725; also Cornell
                 University}
\altaffiltext{6}{NASA Ames Research Center, Moffett Field, CA 94035}
\altaffiltext{7}{Department of Physics and Astronomy, Johns Hopkins University,
                 3400 North Charles Street, Baltimore, MD 21218}
\altaffiltext{8}{I. Physikalisches Institut, Universit\"{a}t zu K\"{o}ln,
                 Z\"{u}lpicher Strasse 77, D-50937 K\"{o}ln, Germany}

\begin{abstract}

We present a 17-point map of the M17SW cloud core in the
$1_{10}\rightarrow1_{01}$ transition of ortho-\water\ at 557 GHz
obtained with \swas.  Water emission was detected in 11 of the 17
observed positions.  The line widths of the \water\ emission vary
between 4 and 9 km s$^{-1}$, and are similar to other emission lines
that arise in the M17SW core.  A direct comparison is made between the
spatial extent of the \water\ emission and the \thco\ $J = 5\rightarrow4$
emission; the good agreement suggests that the \water\ emission arises
in the same warm, dense gas as the \thco\ emission.  A spectrum of
the \wateriso\ line was also obtained at the center position of the
cloud core, but no emission was detected.  We estimate that the average
abundance of ortho-\water\ relative to \hh\ within the M17 dense core is
approximately $1\times10^{-9}$, 30 times smaller than the average for
the Orion core.  Toward the \hii\ region/molecular cloud interface in
M17SW the ortho-\water\ abundance may be about 5 times larger than in
the dense core.

\end{abstract}


\section{Introduction}

M17SW is a prototypical giant molecular cloud core at a distance of
approximately 2.2 kpc \citep*{chi80} that was first studied in detail by
\citet{lad76}.  The dense core lies adjacent to a large optical \hii\
region and is oriented such that the \hii\ region/molecular cloud
interface is viewed nearly edge-on. Because of this favorable geometry,
this region has been the subject of numerous investigations studying
the effect of UV radiation on heating, dissociation, and ionization of
the gas within the molecular cloud.  The detection of emission from
neutral and ionized fine structure lines of atomic carbon and high-J CO
lines \citep{kee85, har87, gen88, stu88, mei92} well away from the \hii\
region/molecular cloud interface suggests that the UV photons penetrate
deep into the M17SW cloud core.  The ability of the UV photons to
penetrate into the cloud has been attributed to the clumpy structure of
the cloud core.  Indirect evidence for a clumpy cloud structure had
been previously suggested by \citet{sne84}, however high resolution
observations of CO and CS presented by \citet{stu90} reveal more
directly the complex structure of this cloud core.

Besides studies of the water maser emission at 22 GHz, only one
attempt has been made to detect water in M17SW. \citet{wat80} report a
marginal detection of the 183 GHz line of water toward the dense core.
With the {\it Submillimeter Wave Astronomy Satellite} (\swas) we have
observed the lowest energy rotational transition of ortho-\water\ and
ortho-\wateriso.  These transitions have upper state energies only 27
K above the ortho-water ground state, making observations with \swas\ a
powerful means to probe water in the warm, dense molecular gas of M17SW.
In this Letter we present a 17-point map that reveals extended \water\
emission from this region.  Observations of the \wateriso\ transition were
made toward the center of the cloud core.  Based on these observations
we estimate the relative abundance of ortho-water in M17SW and compare
the results with that found for the extended water emission detected in
Orion by \swas\ \citep{sne00}.

\begin{figure*}[t]
\epsscale{1.0}
\plotone{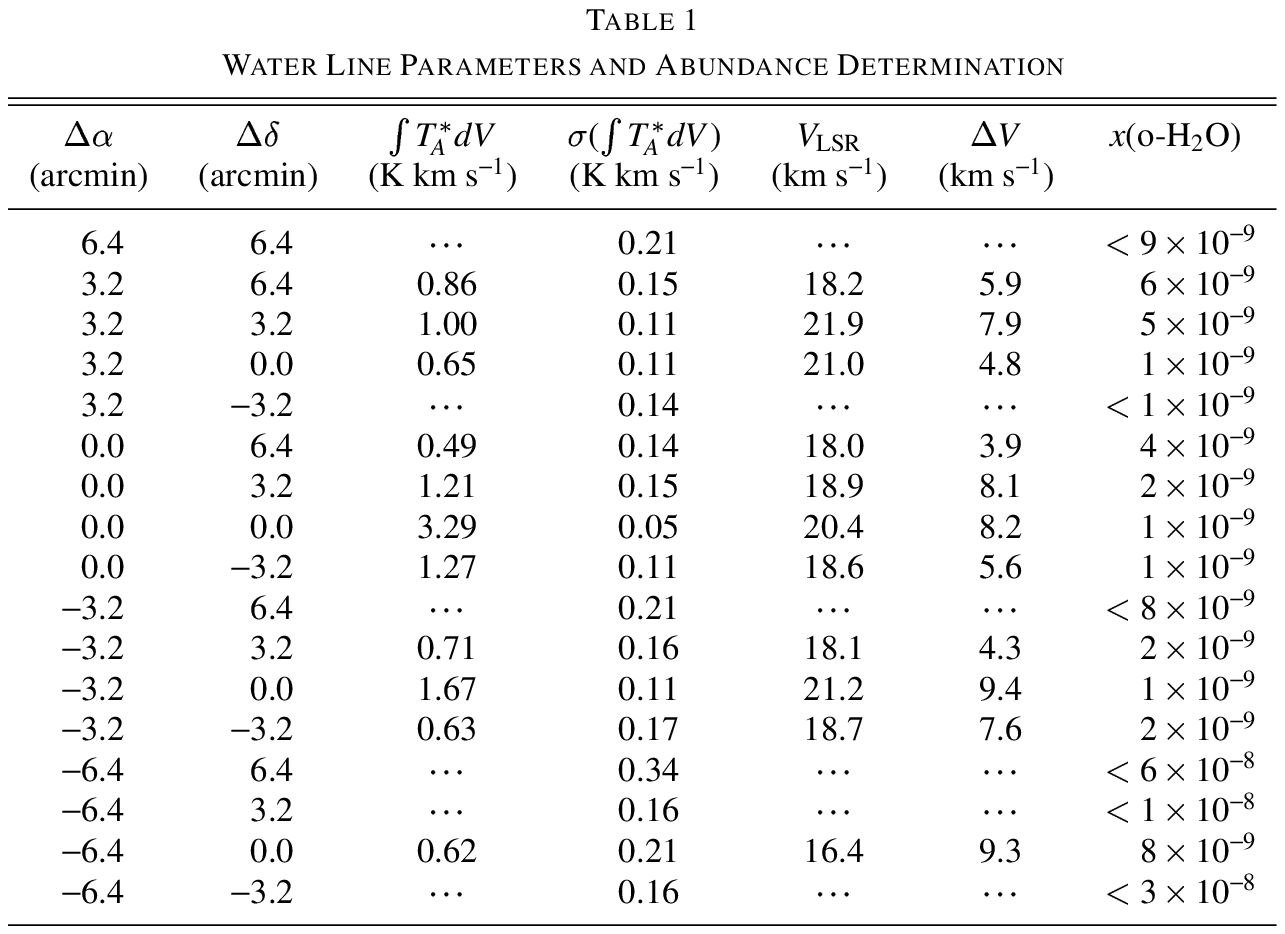}
\end{figure*}

\begin{figure*}[b]
\epsscale{1.10}
\plotone{m17_h2o_fig1.ps}
\caption{Spectra of the $1_{10}\rightarrow1_{01}$ transition of
ortho-\water\ ({\it heavy lines}) and the $J=5\rightarrow4$ transition
of \thco\ ({\it light lines}) obtained in M17SW.  The \thco\ spectra are
divided by a factor of 10 in all cases except at positions (0.0, 0.0) and
(-3.2, 0.0) where the spectra are divided by 20.  The sixteen spectra
of each transition make up a $4\times4$ map obtained on a regular grid
separated by 3\farcm2.  The spectra are shown in their correct relative
positions on the sky.  Offsets in arcminutes relative to position $\alpha
= 18^{\rm h} 20^{\rm m} 22\fs1$, $\delta = -16\arcdeg 12\arcmin 37\arcsec$
(J2000) are indicated on each spectrum.
\label{fig1}}
\end{figure*}

\section{Observations and Results}

The observations of \water\ in M17SW were obtained by \swas\ during the
period 1999 March--1999 June and the observations of \wateriso\ were
obtained during 1999 August--1999 October.  The data were acquired by
nodding the satellite alternatively between M17 and a reference position
free of molecular emission.  Details concerning data acquisition,
calibration, and reduction with \swas\ are presented in \citet{mel00}.
Observations of the $1_{10} \rightarrow 1_{ 01}$ transition of \water\
at a frequency of 556.936 GHz were obtained at 17 positions in the cloud.
The offsets of the 17 spectra relative to position $\alpha = 18^{\rm h}
20^{\rm m} 22\fs1$, $\delta = -16\arcdeg 12\arcmin 37\arcsec$ (J2000)
are given in Table 1.  The absolute pointing accuracy of \swas\ is better
than 5\arcsec\ \citep{mel00}.  Integration times for these observations
were typically 3--7 hr per position, except the center position
which had an integration time of 80 hr.  In the opposite receiver
sideband of \water, we simulataneously obtained spectra of the \thco\
$J = 5 \rightarrow\ 4$ transition.  The center position in M17SW was
also observed in \wateriso\ at a frequency of 547.676 GHz for nearly
100 hr.  The \swas\ beam is elliptical, and at the frequency of the
water transitions has angular dimensions of $3\farcm3 \times 4\farcm5$.
The data shown in this Letter are not corrected for the measured \swas\
main beam efficiency of 0.90.

We also made use of the Five College Radio Astronomy Observatory (FCRAO)
14~m telescope to acquire data for our analysis.  A map of a $12\arcmin
\times 12\arcmin$ region was obtained in the $J=1 \rightarrow 0$
transition of \thco\ and combined with the lower signal-to-noise ratio
map of \citet*{wil99}, and a map of a $6\arcmin \times 6\arcmin$ region
was obtained in the $J = 6 \rightarrow 5\ (K = 0,1,2,3,4)$ transitions of
\chhhcch.  These data are used to provide an estimate of the temperature
and column density of the gas for our analysis of the water emission.

Sixteen of the seventeen \water\ spectra were obtained on a 4 by 4 grid
with a spacing of 3\farcm2.  These spectra, shown in their relative
positions on the sky, are presented in Figure 1.  Also shown in this
figure are spectra of the $J = 5 \rightarrow\ 4$ transition of \thco.
A much larger map of the $J = 5 \rightarrow\ 4$ transition of \thco\
obtained by \swas\ is presented in \citet{how00}.  The strongest \water\
emission of 0.38 K was detected toward the map reference position
located at the center of the dense core.  The seventeenth spectrum was
obtained 6\farcm4 east and 6\farcm4 north of the reference position,
but no water emission was detected.  For the positions where \water\ was
detected, we fit the emission with a Gaussian line shape; the results of
this fitting are presented in Table 1.  The emission at the center of
M17SW is about 5 times weaker than the narrow component of the \water\
emission detected in Orion by \swas\ \citep{sne00}.  The \water\ FWHM line
widths vary between 4 and 9 km s$^{-1}$ and are very similar to those
found for CS and CO emission in this region \citep{lad76,sne84,stu90}.
Although the \water\ spectra appear singly peaked, comparison with \thco\
$J=5\rightarrow4$ spectra (see Fig.\ 1) indicates that in some directions
the \water\ emission may be slightly self-absorbed.  When differences
in angular resolution are taken into account, the spatial extent of the
\water\ emission is similar to that found in CS and the lower rotational
transitions of \ceio\ and \thco\ \citep{sne84, stu90, wil99}.  A more
direct comparison of the spatial distributions can be made between the
\water\ emission and the \thco\ $J = 5 \rightarrow\ 4$ emission obtained
simultaneously with \swas.  The results shown in Figure 1 reveal a very
strong correlation between the emission in these two lines.  The gas
probed by the \thco\ $J = 5 \rightarrow\ 4$ emission has temperatures in
the range 30 to 60 K and densities $>1\times10^5 cm^{-3}$ \citep{how00}.
The similarity in emission properties suggests that the \water\ emission
arises from the same warm, dense gas that gives rise to \thco\ emission
as well as the emission in CS and the lower rotational transitions of CO.

Figure 2 shows an expanded view of the \water\ spectra obtained toward
the center of the M17SW core along with a spectrum of \wateriso\
obtained in the same direction.  No \wateriso\ emission was detected.
Using the same line width and line center velocity determined from a
gaussian fit to the \water\ line at the center, we fit the \wateriso\
line and set a 3$\sigma$ upper limit on the integrated intensity, $\int
T_A^* dV < 0.15$ K km s$^{-1}$.  Thus the integrated intensity ratio of
H$_2$O/H$_2^{18}$O is $>$ 22 (3$\sigma$).

\section{Analysis}

\citet{sne00} present an analytical expression for the relative
abundance of ortho-water in the low-collision rate or effectively thin
limit based on the work of \citet{lin77}.  For large optical depths,
the low-collision rate limit is satisfied if $C\tau_o/A << 1$, where
$C$ is the collisional de-excitation rate coefficient, $\tau_o$ is the
line center optical depth, and $A$ is the spontaneous emission rate.
In this limit, the water integrated intensity increases linearly with
increasing water column density.  \citet{lin77} showed that this limit
will be met if the main beam antenna temperature is sufficiently weak,
and satisfies

\begin{equation}
T_{\rm mb} \ll\ {h\nu\over4k} {\rm exp}(-h\nu/kT_K).
\end{equation}

\noindent
In this limit the line intensity is proportional to the column
density of water irrespective of the line optical depth.  At a kinetic
temperature of 40 K, the 557 GHz line of water is effectively thin if
the antenna temperature is less than 3.4 K.  The maximum observed
intensity for the \water\ emission in M17SW is 0.42 K, after
correction for the \swas\ main beam efficiency.  Therefore, unless the
area filling factor for the \water\ emission is much smaller than 0.1,
the \water\ emission in M17SW is effectively optically thin.

Based on the simple analytical expression, we can estimate the water
abundance in the M17SW core.  The M17SW core has a density of
approximately $6\times10^{5}$ cm$^{-3}$ \citep{sne84, wan93}, and based
on our FCRAO \thco\ observations, the core has an \hh\ column density
of $2\times10^{23}$ cm$^{-2}$.  The kinetic temperature of the core 
is estimated to be 50 K based on the analysis of our \chhhcch\
observations.  Using the main-beam--corrected integrated intensity of
the water emission to-
\hbox to 90.0mm{
ward the center of M17SW of 3.6 K km s$^{-1}$ we estimate the
}

\centerline{\includegraphics[width=0.92\hsize,clip]{m17_h2o_fig2.ps}}
\figcaption{Spectra of \water\ ({\it top}) and \wateriso\ ({\it bottom})
obtained with \swas\ toward the center of the M17 cloud core at position
$\alpha = 18^{\rm h} 20^{\rm m} 22\fs1$, $\delta = -16\arcdeg 12\arcmin
37\arcsec$ (J2000).
\label{fig2}}
\vspace{0.9\baselineskip}

\noindent
beam average relative abundance of ortho-\water\ to be
$8\times10^{-10}$.  Since the dense core in M17SW is large and fills
most of the \swas\ main beam, this abundance estimate should be
reasonably accurate.

The determination of the abundance of ortho-\water\ throughout our map
requires a more detailed model of the temperature, density, column
density, and velocity dispersion of the gas in M17SW.  Analysis of our
\chhhcch\ data yields temperatures of 50 K toward the center of the
core, decreasing to 30 K where the \chhhcch\ emission became too weak
to detect.  These temperatures agree well with those found by
\citet{ber94} and \citet{how00}.  Beyond the point where \chhhcch\ is
detectable, we use the kinetic temperatures derived in \citet{how00},
which were generally in the range of 25 to 45 K. The density studies of
\citet{sne84}, \citet{wan93}, and \citet*{ber96} all indicate that the
density across the M17SW core is relatively uniform and has a value of
$6\times10^{5}$ cm$^{-3}$.  The dense core is elongated north/south
with a long axis of approximately 6\arcmin\ and a short axis of
approximately 4\arcmin. Beyond the core, we assume the density is
$1\times10^5$ cm$^{-3}$ based on the analysis of \citet{wil99} and
\citet{how00}.  The gas column density was derived from the \thco\ data
assuming LTE and a \thco/\hh\ abundance ratio of $1.5\times10^{-6}$.
The velocity dispersion of the gas along each line of sight was
determined from the \thco\ line width.

We modeled the \water\ emission identically to that described in
\citet{sne00} for the Orion cloud core.  We used both the para- and
ortho-\hh\ collision rates with ortho-\water\ \citep*{phi96} and assumed
that the ratio of ortho- to para-\hh\ is in LTE (which at 40 K implies
an ortho-to-para ratio of 0.1).  As in the Orion analysis, we do not
include the continuum emission from dust; however, \citet{sne00} argue
that this will have only a minor impact on the derived water abundance.
The 5 lowest levels of ortho-water are included in our calculations.
We define the physical properties of M17SW on a 44 arcsec grid, much
smaller than the \swas\ resolution.  We proceed by assuming a water
abundance, and then compute the emission that would be predicted within
the \swas\ beam.  We then vary the \water\ abundance until the predicted
emission agrees with observations.  Thus, for each of the 17 positions we
determine the best average \water\ abundance for the gas that contributes
to that \swas\ observation.  For positions with no detections, we used
the 3$\sigma$ upper limit on the integrated intensity to establish a
limit on the \water\ abundance.  We note that where \water\ was detected,
our model predicts that the emission is optically thick, but effectively
thin, although we do not make either assumption in our model.

The results of our abundance analysis are presented in Table 1.
With the exception of the region northeast of the core center, the
abundance of ortho-\water\ relative to \hh\ is approximately constant
with values between 1 and 2$\times10^{-9}$ and consistent with our
estimates based on the analytical expression.  However, in the three
positions in the northeast corner of the map shown in Figure 1,
the relative abundance of ortho-\water\ is roughly a factor of 5
larger than that found toward the core.  The region of enhanced \water\
abundance lies toward the interface between the \hii\ region and the
molecular cloud \citep*{fel80}.  However since the physical conditions,
particularly density and temperature, are not as well known for this
region, the abundance determination is more uncertain.

We have also determined the abundance of \water\ based solely on our
observations of \wateriso\ toward the core center.  Using the same
model described above, we derive a $3\sigma$ upper limit on the
ortho-\wateriso\ abundance of $6\times10^{-11}$.  Assuming a ratio of
H$_2$O/H$_2^{18}$O of 500, provides a $3\sigma$ upper limit of
$3\times10^{-8}$ for the relative abundance of ortho-\water.  If our
model of the \water\ line is correct, then the \wateriso\ line should
have an integrated intensity about 20 times smaller than our $3\sigma$
upper limit, making this line nearly impossible to detect with \swas.

Our modeling has ignored several potentially important effects that
were discussed in \citet{sne00}.  The most important effects are line
scattering by an extended halo that might surround the M17SW core, and
cloud structure on angular scales much smaller than 44\arcsec.  The fact
that the emission lines of \water\ are not strongly self-absorbed and
that there is good agreement between the spatial extent of the \water\
emission and that of optically thin tracers of the dense core, provide
strong arguments against line scattering being significant in this source.
Even if scattering were important for \water, the optical depth in
the \wateriso\ line would be 500 times smaller and the effect of line
scattering (in \wateriso) would be negligible.

The M17SW core is known to have substantial structure on a variety
of angular scales.  We have assumed that on angular scales less than
44\arcsec\ that the the area filling factor of \water-emitting gas
is near unity.  If the area filling factor is instead very small, our
modeling will underestimate the optical depth of the \water\ emission and
consequently the importance of collisional de-excitation.  However, unless
the area filling factor is less than 0.1, the presence of cloud structure
will not impact our determination of the \water\ abundance.  \citet{stu90}
estimated that the area filling factor of the gas within M17SW to
be greater than unity, well above the limit that would cause concern.
However, regardless of the impact of scattering and unresolved structure
on the \water\ analysis, these effects will be unimportant for \wateriso\,
and the non-detection of \wateriso\ implies that the fractional abundance
of ortho-\water\ cannot be greater then $3\times10^{-8}$.

\section{Discussion and Summary}

\swas\ has made the first detection of thermal water emission from
M17 SW.  The emission observed by \swas\ is consistent in line
width, line velocity, and spatial extent with the emission arising
predominately from the warm dense core gas.  The \swas\ observations
allow us to make the first estimate of the water abundance in this
well-studied core.  The average abundance of ortho-\water\ relative to
\hh\ in the cloud core is (1--2)~$\times10^{-9}$.  Northeast of the core,
toward the \hii\ region/molecular cloud interface, the relative water
abundance is approximately five times larger.  Based on the \wateriso\
spectrum, the $3\sigma$ upper limit on the relative \water\ abundance
is $3\times10^{-8}$.  Uncertainties in the derived abundance of water
are dominated by the uncertainties in the physical conditions primarily
density.  In the effectively thin limit, the abundance of \water\ is
inversely proportional to the density \citep[see][]{sne00}.  Therefore,
if we had assumed a smaller average density, we would have derived
a proportionally higher water abundance.  The study of \citet{ber96}
concludes that the bulk of the column density toward the M17 core arises
in the dense gas.  Thus, we believe that uncertainties in the density
and density structure cannot conspire to increase the water abundance
by more than an order of magnitude.

The average relative abundance of water in the M17SW core is 30 times
smaller than the average in Orion \citep{sne00}.  Although there is
substantial variation in the \water\ abundance in Orion, the abundance is
always significantly larger than in the M17SW.  Only near the interface
with the \hii\ region in M17SW does the abundance of \water\ approach
values found for Orion.  The enhanced abundance of water in this interface
region could be a result of the evaporation of water-ice--rich mantles
from interstellar grains exposed to radiation from the \hii\ region.
Further discussion of the chemical implications of these \water\ abundance
results is presented in \citet{ber00}.

\acknowledgements

This work was supported by NASA \swas\ contract NAS5-30702 and NSF
grant AST 97-25951 to the Five College Radio Astronomy Observatory.
R. Schieder and G. Winnewisser would like to acknowledge the generous
support provided by the DLR through grants 50 0090 090 and 50 0099 011.

\end{document}